\documentclass[12pt]{article}

\usepackage[margin=1in]{geometry}
\usepackage[utf8]{inputenc}
\usepackage{amsmath,amssymb,amsthm}
\usepackage{mathtools}
\usepackage{bm}
\usepackage{graphicx}
\usepackage{xcolor}
\usepackage[colorlinks=true, linkcolor=blue, citecolor=blue, urlcolor=blue, hypertexnames=false]{hyperref}
\hypersetup{pdftitle={A hinged honeycomb with zero bulk modulus retaining more than four-fifths of its constituent's shear modulus}, pdfauthor={Chun-Teh Chen}}
\usepackage[numbers]{natbib}
\usepackage{microtype}
\usepackage{booktabs}
\usepackage{caption}
\usepackage{float}

\newcommand{\dev}{\operatorname{dev}}
\newcommand{\tr}{\operatorname{tr}}
\newcommand{\dvg}{\operatorname{div}}
\newcommand{\conv}{\operatorname{conv}}

\newtheoremstyle{result}{\topsep}{\topsep}{\normalfont}{}{\bfseries}{.}{ }{\thmname{#1}\thmnumber{ #2}\thmnote{ (#3)}}
\theoremstyle{result}
\newtheorem{result}{Result}

\graphicspath{{./}{figures/}}
\allowdisplaybreaks

\title{\textbf{A hinged honeycomb with zero bulk modulus retaining more than four-fifths of its constituent's shear modulus}}

\author{Chun-Teh Chen}

\date{
  \small Department of Materials Science and Engineering, University of California,\\
  \small Berkeley, CA, USA\\[4pt]
  \small Correspondence: \texttt{chunteh@berkeley.edu}}

\begin{document}

\maketitle

\begin{abstract}

A Poisson's ratio near $-1$ indicates only that the bulk modulus is small compared with the shear modulus. In solid--void structures, the mechanism that frees the dilation usually weakens the resistance to shear as well, resulting in both moduli becoming small. In two-dimensional linear elasticity, we demonstrate that this loss is not inevitable. We consider a honeycomb of regular hexagonal blocks of a single incompressible isotropic elastic solid, joined along their whole edges by ideal interfaces that allow relative sliding along a direction inclined to the edge normal. Collective infinitesimal block rotations produce an exact dilational mechanism, so the effective bulk modulus vanishes, while sixfold symmetry ensures isotropy. The interfaces, however, transmit traction along their whole length. In a solid--void realization, each interface is replaced with fine solid plates separated by void, which bend easily yet retain their capacity to transmit axial force. Such dilational materials, which expand or contract freely while resisting every change of shape, could serve as interlayers that accommodate thermal or swelling mismatch while still transmitting shear, and as components in stents and deployable structures that change size without changing shape. An explicit, statically admissible stress field and the complementary energy principle yield a rigorous lower bound on the retained shear modulus, and a limiting argument transfers this bound to the solid--void mixtures. The honeycomb retains more than four-fifths of its constituent's shear modulus at zero bulk modulus: the supremum $S$ of the normalized shear modulus of such mixtures satisfies $S>0.8528>4/5$, exceeding the value attained by Milton's construction.
\end{abstract}

\section{Introduction}\label{sec:intro}

Most common materials become narrower when stretched. Auxetic materials instead expand laterally, yielding a negative Poisson's ratio \citep{Evans1991}. In isotropic linear elasticity, this response is governed by the relative resistance to changes in volume and shape \citep{Greaves2011}. In two dimensions, Poisson's ratio is $\nu=(\kappa-\mu)/(\kappa+\mu)$, where $\kappa$ is the planar bulk modulus and $\mu$ is the shear modulus. Since $\kappa,\mu\ge0$, $\nu$ lies between $-1$ and $1$. Most solids resist changes of volume more strongly than changes of shape, so $\kappa>\mu$ and $\nu>0$; a negative Poisson's ratio requires $\kappa<\mu$, and the limit $\nu\to-1$ corresponds to $\kappa/\mu\to0$.

Geometry offers several paths to this response. The inward-pointing ribs of a re-entrant foam unfold under tension \citep{Lakes1987}; rigid squares hinged at their corners rotate alternately clockwise and anticlockwise, opening the gaps between them \citep{Grima2000}; and chiral honeycombs expand through coupled ligament bending and node rotation \citep{Prall1997}. Such mechanisms can make dilation much easier than other forms of deformation. Materials that expand or contract freely yet resist every change of shape are called dilational \citep{Milton1992,Buckmann2014}. However, approaching $\nu=-1$ fixes only the ratio $\kappa/\mu$ and does not constrain $\mu$ itself, which may become small as well \citep{Milton2021open}. In solid--void structures, the hinges are thin ligaments of the solid, and thinning them to free the dilation usually weakens their resistance to shear. The objective is therefore to make the bulk modulus vanish while preserving a large fraction of the constituent's shear modulus. Such materials would serve, for instance, as interlayers that accommodate thermal or swelling mismatch while still transmitting shear, and as components in stents and deployable structures that change size without changing shape.

\citet{Milton2021open} posed a precise version of this question. Consider a two-dimensional isotropic material composed of one isotropic, incompressible elastic solid of shear modulus $\mu$ and void. Among such materials with zero effective bulk modulus, how large can the ratio $\mu^{*}/\mu$ be? He established an attainable limiting ratio of $4/5$ and asked whether it is optimal. His construction first forms an anisotropic ``crystal'' of solid slabs separated by thin layers that are compliant normal to the slabs, then forms an optimal isotropic polycrystal of that crystal \citep{Avellaneda1996,Milton2021planar}. To our knowledge, $4/5$ is the largest value rigorously established before the present work.

\begin{figure}[t]
  \centering
  \includegraphics[width=\textwidth]{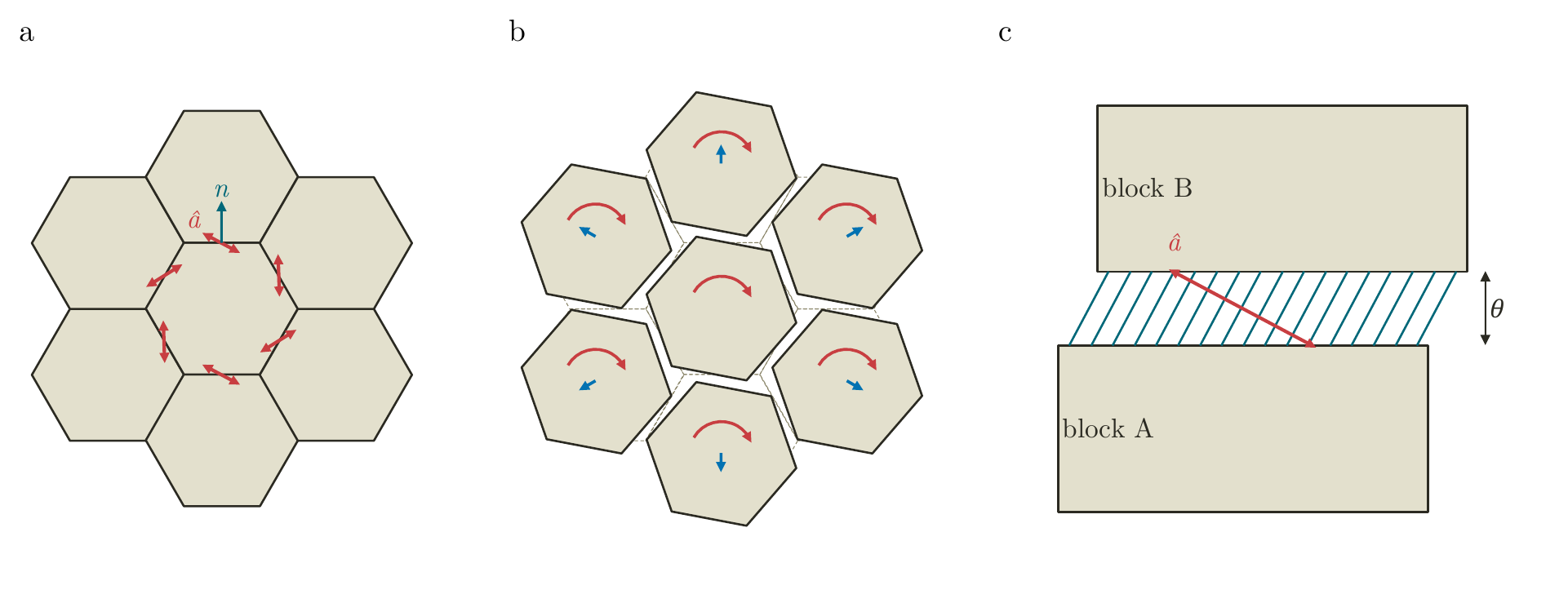}
  \caption{\textbf{The ideal honeycomb and its solid--void realization.} (a) Regular hexagonal blocks joined along their edges by sliding interfaces. The free displacement direction $\hat{a}$ makes an angle $\gamma$ with the edge normal $n$. (b) Schematic of the infinitesimal dilational mechanism: all blocks rotate by the same amount (red arrows) while their centers move outward (blue arrows), producing macroscopic dilation without straining the blocks. The displacements are exaggerated. (c) An interface replaced with a strip of solid plates separated by void. The plates run perpendicular to $\hat{a}$, and $\theta$ denotes the normal separation of the block faces. The geometry is illustrated at $\gamma=62^{\circ}$ in (a) and (c).}
  \label{fig:1}
\end{figure}

Here, we demonstrate that $4/5$ can be exceeded by a honeycomb of regular hexagonal blocks joined along their whole edges by sliding interfaces (Fig.~\ref{fig:1}). Each interface allows relative displacement along a direction tilted from the edge normal by an angle $\gamma$, while transmitting traction perpendicular to that direction. When all hexagons rotate together, their collective motion produces uniform macroscopic dilation without straining the blocks. Therefore, the effective bulk modulus vanishes, and sixfold symmetry makes the effective elasticity tensor isotropic. The interfaces, however, still transmit distributed tractions along the edges. An explicit stress field allows us to quantify the shear modulus retained by this construction.

To state the result precisely, we identify a symmetric $2\times2$ tensor $A$ with its Mandel vector $(A_{11},A_{22},\sqrt{2}A_{12})\in\mathbb{R}^{3}$, and a fourth-order elasticity tensor with the symmetric $3\times3$ matrix acting on these vectors. The same symbol denotes both a tensor and its Mandel representation. This identification preserves inner products, so $A\cdot B=A:B$. Let $h=(1,1,0)/\sqrt{2}$, corresponding to $I/\sqrt{2}$, and let $P_{\dev}=I_{3}-h\otimes h$ be the orthogonal projection onto the deviatoric plane $h^{\perp}$. An isotropic elasticity tensor has the form
\begin{equation}\label{eq:1}
  C=2\kappa^{*}\, h\otimes h+2\mu^{*}P_{\dev},
\end{equation}
so $\kappa^{*}=0$ is equivalent to $Ch=0$.

Let $\mathcal{G}$ be the closure, in the space of elasticity tensors, of the finite effective tensors of periodic mixtures of the fixed incompressible solid and void, in arbitrary proportions. Milton's question concerns
\begin{equation}\label{eq:2}
  S=\sup\left\{ \frac{\mu^{*}}{\mu}:C^{*}\in\mathcal{G},\quad C^{*}\: \text{isotropic},\quad \kappa^{*}=0 \right\}.
\end{equation}
The closure includes the limits of hierarchical constructions used to obtain Milton's value. Together with the uniform-strain upper bound $\mu^{*}\le\mu$ \citep{Hill1963}, his construction establishes  $4/5\le S\le1$. We prove the improved lower bound $S\ge P/Q>4/5$, where $P/Q=0.8528586652\ldots$ is the exact rational value certified in the present work. Results for compressible constituents can be related to this reference problem through the Cherkaev--Lurie--Milton compliance transformation, with corresponding adjustments to the normalized bounds \citep{Cherkaev1992}.

Several earlier constructions share features of this geometry. \citet{Bacigalupo2020} studied rigid hexagonal blocks connected by elastic interfaces, including those represented by parallel beams, while \citet{Ravirala2007} studied a rigid-particle model of sliding, interlocking hexagons. \citet{Acuna2022} obtained zero bulk modulus with finite shear stiffness in ideal models of rotating units coupled by springs. These studies do not establish a shear modulus exceeding four-fifths of that of the constituent solid at zero bulk modulus for a mixture of one elastic solid and void. Building on \citet{Sigmund2000}, \citet{Ostanin2018} numerically explored related isotropic architectures with solid regions connected by inclined laminates but did not establish the attainable shear modulus at zero effective bulk modulus. \citet{Milton2017} characterized possible effective tensors of solid--void mixtures with variational bounds and multiscale constructions of the kind used here but left the largest shear modulus at zero bulk modulus open.

The proof has three steps. We first establish the isotropy and exact dilational mechanism of the ideal honeycomb for $0\le\gamma<90^{\circ}$. At the fixed angle $\tan\gamma=100/\sqrt{3}$ ($\gamma \approx 89.01^{\circ}$), we then use a single statically admissible stress field and the complementary energy principle to prove $\mu^{*}/\mu\ge P/Q$. Lastly, we replace each sliding interface with fine plates of the constituent solid separated by void. Increasing the number of plates at a fixed interface thickness yields limiting tensors in $\mathcal{G}$ with zero bulk modulus; then reducing the interface thickness makes the lower bound on their normalized shear modulus approach $P/Q$.

\section{Results}\label{sec:results}

\subsection{The hinged honeycomb}\label{sec:2.1}

An ideal sliding interface joins two elastic blocks $A$ and $B$ while allowing relative displacement along a unit direction $\hat{a}$. Displacement perpendicular to $\hat{a}$ is constrained, and sliding along $\hat{a}$ is frictionless, so the interface transmits traction only along $\hat{t}=\hat{a}^{\perp}$. Let $n$ be the unit normal pointing from block $A$ to block $B$, and define $[u]=u_{B}-u_{A}$, where $u_{A}$ and $u_{B}$ are the displacement traces at the same interface point. Writing $\sigma$ for the Cauchy stress and using the same normal $n$ on both sides, the interface conditions are
\begin{equation}\label{eq:3}
  [u]\parallel\hat{a},\qquad \sigma n\: \text{continuous},\qquad (\sigma n)\cdot\hat{a}=0.
\end{equation}
We consider a periodic honeycomb of regular hexagonal blocks $B_{j}$, each with edge length $1$, made from the incompressible solid with shear modulus $\mu$. Adjacent blocks are joined along their whole edges by sliding interfaces. We choose one complete hexagon as the periodic cell $\Omega$. All other blocks are translates of $\Omega$ by vectors $r$ in the lattice generated by the center-to-center vectors $dn$, where $d=\sqrt{3}$ is the center spacing. For a prescribed symmetric macroscopic strain $E$, the effective elasticity tensor is defined as
\begin{equation}\label{eq:4}
  E\cdot C^{*}E=\min_{u}\frac{1}{|\Omega|}\int_{\Omega} 2\mu\, |\epsilon(u)|^{2}\, dX,
\end{equation}
where $\epsilon(u)=1/2(\nabla u+\nabla u^{\mathsf{T}})$. The admissible fields satisfy $u(X)=EX+\tilde{u}(X)$, with $\tilde{u}$ periodic, $u|_{\Omega}\in H^{1}(\Omega;\mathbb{R}^{2})$, and $\dvg u=0$ in $\Omega$. The field on each translated block is given by $u(X+r)=u(X)+Er$. Across every interface, the displacement traces satisfy $[u]\cdot\hat{t}=0$. Therefore, displacement jumps are allowed along $\hat{a}$, and the interfaces store no energy. A macroscopic mechanism is an admissible field with zero energy and nonzero macroscopic strain $E$. Its existence implies $C^{*}E=0$, since $C^{*}$ is positive semidefinite.

For the hinged honeycomb $H_{\gamma}$, the interface directions are
\begin{equation}\label{eq:5}
  \hat{a}=\cos\gamma\, n+\sin\gamma\, \tau,\qquad \hat{t}=\cos\gamma\, \tau-\sin\gamma\, n,\qquad \gamma\in[0^{\circ},90^{\circ}),
\end{equation}
where $n$ is the outward normal of a hexagon edge and $\tau=\hat{z}\times n$ is its counterclockwise tangent. The rule is independent of which neighboring block is used: reversing $n$ and $\tau$ reverses both $\hat{a}$ and $\hat{t}$, leaving their directions and the interface conditions unchanged. As $\gamma$ approaches $90^{\circ}$, the free displacement becomes nearly tangential to the edge, while the allowed traction becomes nearly normal to it.

To construct a dilational mechanism, we displace each block center $c_{j}$ by $ec_{j}$ and give every block the same infinitesimal rotation of amplitude $\omega$:
\begin{equation}\label{eq:6}
  u_{j}(X)=e\, c_{j}+\omega J(X-c_{j}),\qquad J=\begin{pmatrix} 0 & -1 \\ 1 & 0 \end{pmatrix}.
\end{equation}
Each block has a displacement gradient $\omega J$, so its strain and divergence vanish. The interface constraint remains to be imposed.

For neighboring blocks $A$ and $B$, let $n$ point from $A$ to $B$. Their centers satisfy $c_{B}-c_{A}=dn$. Taking $[u]=u_{B}-u_{A}$ in equation \eqref{eq:6} yields the constant jump $[u]=(eI-\omega J)dn$ along their shared edge. Since $Jn=\tau$, its component in the constrained direction is
\begin{equation}\label{eq:7}
  [u]\cdot\hat{t}=-d(e\sin\gamma+\omega\cos\gamma).
\end{equation}
The same condition applies to every edge and is satisfied by
\begin{equation}\label{eq:8}
  \omega=-e\tan\gamma.
\end{equation}
For every lattice vector $r$, this displacement satisfies $u(X+r)-u(X)=er$, so $u-eX$ is periodic and the macroscopic strain is exactly $eI$, whose Mandel vector is $\sqrt{2}\, e\, h$. With $e\ne0$, \eqref{eq:6} and \eqref{eq:8} therefore yield a macroscopic mechanism and imply $C^{*}h=0$. Adding an affine displacement with a deviatoric gradient to this mechanism produces an admissible field for every symmetric macroscopic strain, so the cell energy in \eqref{eq:4} is finite. At $\gamma=0$, the rotation vanishes, and the mechanism is pure separation of the blocks. The endpoint $\gamma=90^{\circ}$ is excluded because \eqref{eq:7} would require $e=0$.

The honeycomb and the interface rule \eqref{eq:5} are invariant under rotation through $60^{\circ}$ about any block center. This rotation fixes the hydrostatic direction and acts as a rotation through $120^{\circ}$ on the deviatoric plane. Invariance therefore eliminates coupling between hydrostatic and deviatoric strains and forces the symmetric deviatoric block of $C^{*}$ to be a scalar multiple of the identity matrix. Consequently, $C^{*}$ is isotropic, and $C^{*}h=0$ means $\kappa^{*}=0$.

\begin{result}[the honeycomb is isotropic with zero bulk modulus]\label{res:1}
For every $\gamma\in[0^{\circ},90^{\circ})$, the hinged honeycomb $H_{\gamma}$ has the exact infinitesimal dilational mechanism \eqref{eq:6}, \eqref{eq:8}. Its effective elasticity tensor is isotropic and satisfies $\kappa^{*}(H_{\gamma})=0$.
\end{result}

\subsection{A lower bound from a single stress field}\label{sec:2.2}

We obtain a lower bound on the shear modulus by constructing one statically admissible stress field. Such a field $\sigma$ is periodic, symmetric, and divergence-free within each block, has continuous traction across interfaces, and satisfies $(\sigma n)\cdot\hat{a}=0$ on every sliding interface. We write the mean stress as $\Sigma=|\Omega|^{-1}\int_{\Omega} \sigma\, dX$, where $\Omega$ is the complete hexagonal cell. For the incompressible solid, the complementary energy principle yields
\begin{equation}\label{eq:9}
  W_{c}:=\frac{1}{|\Omega|}\int_{\Omega} \frac{|\dev\sigma|^{2}}{4\mu}\, dX\; \ge\; \sup_{E=E^{\mathsf{T}}}\left(\Sigma\cdot E-\frac{1}{2}E\cdot C^{*}E \right),
\end{equation}
where $\dev\sigma=\sigma-1/2(\tr\sigma)I$, and the supremum is over symmetric macroscopic strains.

Result~\ref{res:1} establishes that $C^{*}$ is isotropic and $C^{*}I=0$. Taking $E=eI$ in \eqref{eq:9} yields $e\, \tr\Sigma$ inside the supremum. For finite $W_{c}$, this forces $\tr\Sigma=0$. If $\Sigma\ne0$, finiteness also forces $\mu^{*}>0$; otherwise, the supremum would be unbounded along $E=t\Sigma$. The supremum is then attained at $E=\Sigma/(2\mu^{*})$ and equals $|\Sigma|^{2}/(4\mu^{*})$. Consequently,
\begin{equation}\label{eq:10}
  \frac{\mu^{*}}{\mu}\; \ge\; \frac{|\Sigma|^{2}}{4\mu W_{c}}.
\end{equation}
The numerator divided by $4\mu$ is the complementary energy of the uniform constituent solid under the same mean shear stress. Therefore, for a fixed nonzero $\Sigma$, a smaller trial complementary energy produces a stronger lower bound.

We construct the stress using an Airy potential $\varphi$, with Cartesian components $\sigma_{11}=\varphi_{,22}$, $\sigma_{22}=\varphi_{,11}$ and $\sigma_{12}=-\varphi_{,12}$. These expressions ensure symmetry and equilibrium wherever $\varphi$ is smooth. We divide the hexagon into six triangles by joining its center to its vertices, take $\varphi$ to be a polynomial of degree at most ten on each triangle, and repeat the resulting stress field periodically.

To carry out the calculation in rational arithmetic, we use oblique coordinates $X=\mathsf{p}\, \mathsf{e}_{1}+\mathsf{q}\, \mathsf{e}_{2}$, where $\mathsf{e}_{1}=(1,0)$ and $\mathsf{e}_{2}=(-1/2,\sqrt{3}/2)$. In these coordinates the vertices $v_{0},\ldots,v_{5}$ are $(1,0),(1,1),(0,1),(-1,0),(-1,-1),(0,-1)$, and the metric is $g=\begin{pmatrix} 1 & -1/2 \\ -1/2 & 1 \end{pmatrix}$. Indices are understood modulo six. Let $\varphi_{k}$ denote the polynomial piece on the triangle with oblique-coordinate vertices $0,v_{k},v_{k+1}$. The polynomial pieces satisfy three conditions:
\begin{enumerate}
  \item \textbf{Continuity inside each block.} The potential and its first derivatives agree across each radius. This ensures traction continuity between adjacent triangles.
  \item \textbf{The sliding condition on each edge.} Let $\tau=v_{k+1}-v_{k}$, and let $\nu=(\tau_{\mathsf{q}},-\tau_{\mathsf{p}})$ be the primitive outward integer normal covector. If $H$ is the Hessian of $\varphi$ in oblique coordinates, the condition $(\sigma n)\cdot\hat{a}=0$ becomes
\begin{equation}\label{eq:11}
  H(\tau,\tau)=\lambda H(g^{-1}\nu,\tau),\qquad \lambda=\frac{\sqrt{3}}{2}\tan\gamma.
\end{equation}

Here $H(v,w)=v^{\mathsf{T}}Hw$, with both arguments expressed in oblique coordinates. The physical vector represented by $g^{-1}\nu$ is $2n/\sqrt{3}$.

  \item \textbf{Traction continuity between blocks.} The traction on edge $k$ agrees with that of the neighboring hexagon, whose stress field is translated by $v_{k}+v_{k+1}$ in oblique coordinates.
\end{enumerate}

We choose $\lambda=50$, equivalently $\tan\gamma=100/\sqrt{3}$, so $\gamma\approx89.00771^{\circ}$. Because $\lambda$ is rational, the three conditions are polynomial identities yielding linear equations with rational coefficients for the coefficients of $\varphi$.

We obtain the field by maximizing the right side of \eqref{eq:10} within this polynomial space. The proof requires only verifying the resulting field, regardless of how it is found. After clearing denominators, the coefficients are integers. The field is centrosymmetric, $\varphi_{k+3}(X)=\varphi_{k}(-X)$, so three polynomial pieces determine all six. Exact rational arithmetic verifies all three admissibility conditions, confirms that the mean shear stress is nonzero, and evaluates the quotient in \eqref{eq:10}.

\begin{result}[more than four-fifths of the shear modulus]\label{res:2}
At $\tan\gamma=100/\sqrt{3}$, the stress field generated by a piecewise degree-ten Airy potential certifies
\begin{equation}\label{eq:12}
  \frac{\mu^{*}(H_{\gamma})}{\mu}\; \ge\; \frac{P}{Q}>0.8528>\frac{4}{5}.
\end{equation}
\end{result}
\noindent Here $P/Q$ is the exact rational value of the quotient in \eqref{eq:10} for this field.

\subsection{From sliding interfaces to solid--void mixtures}\label{sec:2.3}

\begin{figure}[t]
  \centering
  \includegraphics[width=\textwidth]{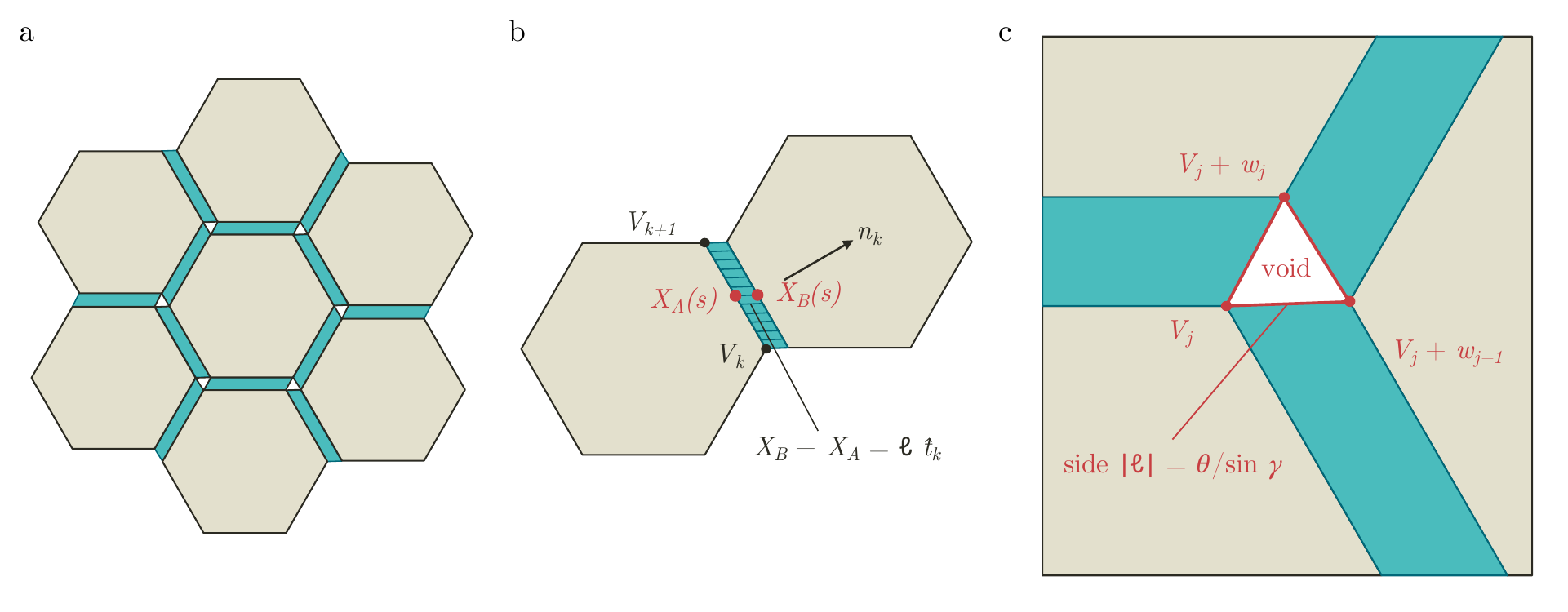}
  \caption{\textbf{Geometry of the thickened tiling.} (a) The structure $H_{\gamma,\theta}$: hexagonal blocks (beige), strip regions (teal), and equilateral void triangles (white) form a periodic tiling. (b) Corresponding points $X_{A}(s)$ and $X_{B}(s)$ on the facing block edges differ by the fixed translation $\ell\hat{t}_{k}$, with $|\ell|=\theta/\sin\gamma$, parallel to the plate direction $\hat{t}_{k}$, allowing each plate to connect both blocks. (c) Enlargement of a three-block junction, showing the equilateral void triangle bounded by the lateral sides of the adjacent strips, with vertices $V_{j}$, $V_{j}+w_{j}$, and $ V_{j}+w_{j-1}$, where $w_{j}$ is the offset vector of Appendix~\ref{sec:A}. The tiling is illustrated at $\gamma=62^{\circ}$ and $\theta=0.16$.}
  \label{fig:2}
\end{figure}

Sections~\ref{sec:2.1} and \ref{sec:2.2} consider ideal sliding interfaces. To realize their behavior using the constituent solid and void, we replace each interface with a strip of solid plates separated by void (Fig.~\ref{fig:1}). Let $\theta$ denote the strip thickness measured normal to the facing block edges. The inclined plates require a tangential offset of magnitude $\theta\cot\gamma$ between these edges. The lattice construction in Appendix~\ref{sec:A} provides this offset and joins each plate to both blocks. Blocks, strips, and equilateral void triangles at the junctions of three blocks form a periodic tiling (Fig.~\ref{fig:2}).

\begin{figure}[t]
  \centering
  \includegraphics[width=\textwidth]{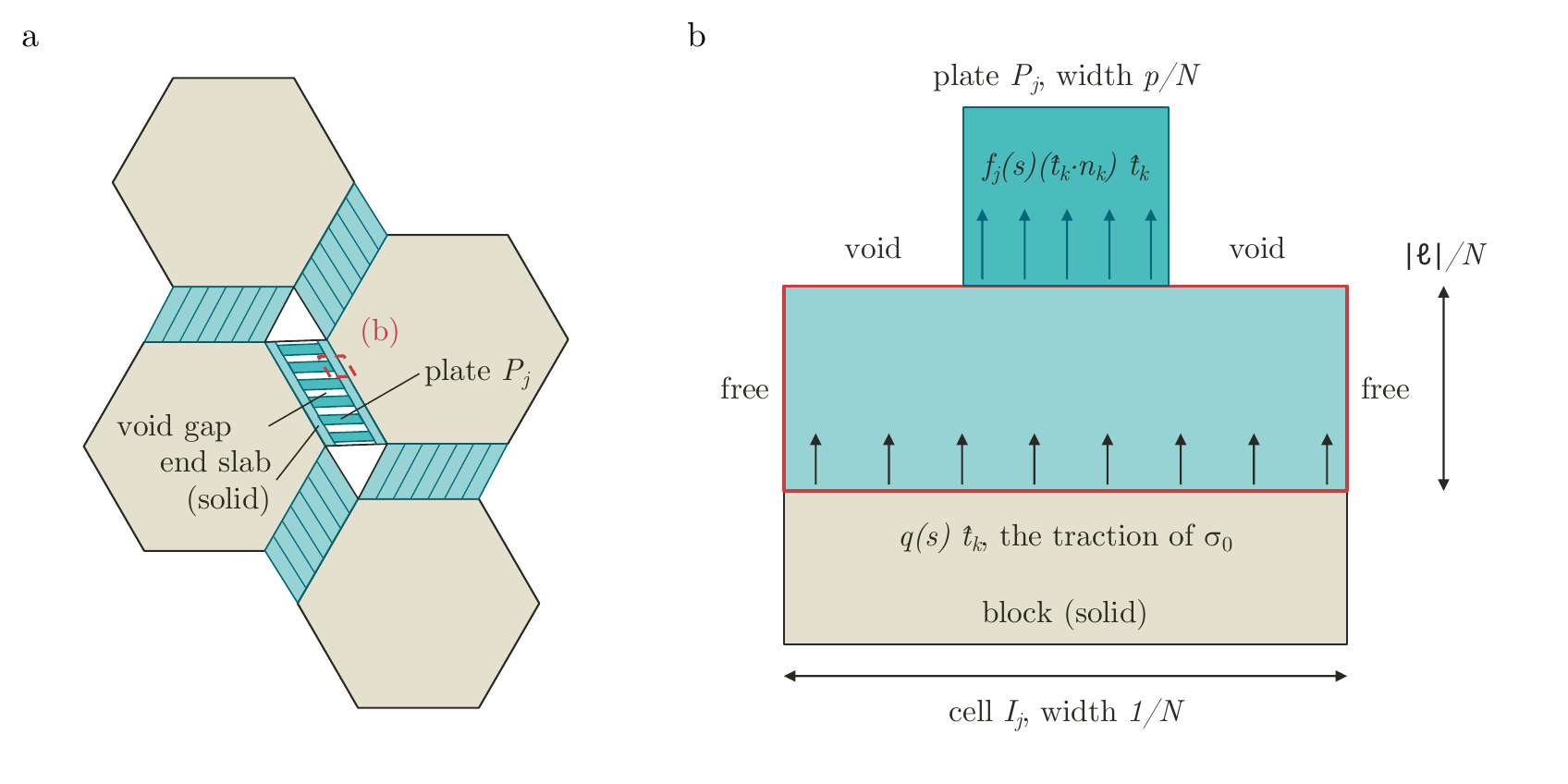}
  \caption{\textbf{Plate structure and transmission of traction.} (a) Detail of the solid--void structure $H_{\gamma,\theta}^{N}$. Each strip contains $N$ solid plates (teal) separated by void and connected to the blocks through thin solid end slabs (light teal), shown in full for one strip and abbreviated in the others. The solid fraction between the end slabs is $p$. The drawing uses $N=6$ and $p=0.55$, with the strip thickness exaggerated for clarity. (b) Schematic of the end-slab cell marked by the dashed box in (a), connecting the block face to a centered plate. The plate's axial trial stress $f_{j}(s)$ varies affinely across its width and is chosen to balance the force and moment supplied by the block-face traction. Arrows show the traction $q(s)\hat{t}_{k}$ from the block face (bottom) and the plate's axial traction (top); the lateral sides and the gaps beside the plate are traction-free, and the cell height is $|\ell|/N$. The stress construction is detailed in Appendix~\ref{sec:B.3}. The widths $1/N$ and $p/N$ are measured along the block face.}
  \label{fig:3}
\end{figure}

Within each strip, $N$ thin solid plates run parallel to $\hat{t}$, separated by void and connected to the blocks by thin solid end slabs (Fig.~\ref{fig:3}). Let $p\in(0,1)$ denote the solid fraction of the strip interior between the end slabs. The resulting solid--void structure $H_{\gamma,\theta}^{N}$ (Appendix~\ref{sec:B.1}) has sixfold symmetry, so its effective elasticity tensor is isotropic.

At fixed $\theta$ and $p$, increasing $N$ makes the plates thinner: they bend more easily while retaining their capacity to transmit axial force.

The argument has two steps. First, at fixed $\theta$, we increase $N$ to obtain limiting tensors with zero bulk modulus and controlled shear stiffness. We then let $\theta\to0$ in the resulting lower bound on the shear modulus.

\textbf{Dilation mode.} The blocks move rigidly as in Section~\ref{sec:2.1}, with the rotation-to-dilation ratio adjusted to the thickened tiling (Appendix~\ref{sec:A.3}), so that facing block points move relative to each other transversely to the plates. The plates accommodate this motion by bending. Appendix~\ref{sec:B.2} constructs an exactly incompressible displacement field matching the blocks and end slabs. Its energy implies $\kappa^{*}(H_{\gamma,\theta}^{N})\le C_{1}N^{-2}$ for sufficiently large $N$, where $C_{1}$ may depend on $\theta$.

\textbf{Shear mode.} We fix $\tan\gamma=100/\sqrt{3}$ and $p=1/2$. Let $\sigma_{0}$ be the stress field of Result~\ref{res:2}, and let $M_{0}$ and $E_{0}$ be its integrated stress and complementary energy over the ideal cell $\Omega$. Its quotient in \eqref{eq:10} is $|M_{0}|^{2}/(4\mu|\Omega|E_{0})=P/Q$.

We build a trial stress for $H_{\gamma,\theta}^{N}$. By \eqref{eq:9}, any statically admissible field provides a lower bound, so the true stress of the new structure is not needed. In the blocks, we take $\sigma_{0}$ itself. Its traction on each block face has the form $\sigma_{0}n=q\hat{t}$, which is exactly the direction the plates can carry. In each plate, choose an axial stress $f\, \hat{t}\otimes\hat{t}$, with $f$ affine across the plate width, to balance the force and moment supplied by the corresponding portion of the block face. An equilibrium stress in each end-slab cell connects the block traction to the plate traction (Fig.~\ref{fig:3}; Appendix~\ref{sec:B.3}).

The plate stresses are bounded by a constant $F$ that depends only on the block-face traction $q$ (Appendix~\ref{sec:B.3}, \eqref{eq:29}), and the plates occupy an area at most $3p\theta$ per periodic cell. The end-slab contributions vanish as $N^{-1}$ at fixed $\theta$. Therefore, the integrated stress $M_{\theta,N}$ and complementary energy $E_{\theta,N}$, taken over a cell $\Omega_{\theta}$ with the stress extended by zero into the void, satisfy
\begin{equation}\label{eq:13}
  \begin{aligned} |M_{\theta,N}-M_{0}| &\le C_{m}\theta+D_{\theta}N^{-1}, \\ E_{\theta,N} &\le E_{0}+C_{e}\theta+D_{\theta}N^{-1}. \end{aligned}
\end{equation}
Here $C_{m}$ and $C_{e}$ are independent of both $\theta$ and $N$; $D_{\theta}$ may depend on $\theta$, but not on $N$.

Restricting the complementary energy principle \eqref{eq:9} to deviatoric macroscopic strains yields
\begin{equation}\label{eq:14}
  \frac{\mu^{*}(H_{\gamma,\theta}^{N})}{\mu}\ge\frac{|\dev M_{\theta,N}|^{2}}{4\mu|\Omega_{\theta}|E_{\theta,N}}.
\end{equation}
This inequality requires no assumption that the trial mean stress is trace-free.

At fixed $\theta$, the bulk estimate and the uniform shear bound $\mu^{*}\le\mu$ limit both moduli of the isotropic effective tensors, so they have subsequential limits as $N\to\infty$. Every such limit is isotropic, belongs to $\mathcal{G}$, and has zero bulk modulus.

For sufficiently small $\theta$, equations \eqref{eq:13}--\eqref{eq:14}, together with $\dev M_{0}=M_{0}$, provide a common positive shear bound for every limit. As $\theta\to0$, the cell area $|\Omega_{\theta}|$ tends to $|\Omega|$, and this bound tends to $P/Q$. Appendix~\ref{sec:B.4} yields the explicit bound and the passage to the limit.

\begin{result}[solid--void mixtures exceed $4/5$]\label{res:3}
The construction above establishes
\begin{equation}\label{eq:15}
  S\ge\frac{P}{Q}>0.8528>\frac{4}{5}.
\end{equation}
\end{result}
\noindent The bound equals that of Result~\ref{res:2} because the corrections in \eqref{eq:13} vanish as $\theta\to0$. Result~\ref{res:3} asserts it for the supremum over solid--void mixtures, which Result~\ref{res:2} alone does not. The estimates also imply that, for sufficiently small fixed $\theta$, all sufficiently large finite $N$ have a normalized shear modulus above $4/5$, while the effective Poisson's ratio tends to $-1$ as $N\to\infty$. The proof concerns homogenized linear moduli and does not specify numerical choices of $\theta$ and $N$.

\section{Conclusions}\label{sec:conclusions}

We prove that $4/5$ can be exceeded in Milton's problem: for two-dimensional isotropic mixtures of an incompressible elastic solid and void with zero effective bulk modulus, $S\ge P/Q>0.8528$. The improvement over Milton's construction arises from the choice of the released deformation mode. Milton's crystal releases a uniaxial strain, and making that mode isotropic by forming a polycrystal costs one fifth of the shear modulus. The hinged honeycomb releases a pure dilation, which is already isotropic, so nothing is lost at that step. The interfaces then transmit traction along their whole length, and with glued interfaces the honeycomb would be the solid itself, so the entire reduction from $\mu$ to $\mu^{*}$ is due to the sliding interfaces alone.

The bound is unlikely to be sharp, because it certifies one structure at one angle with one polynomial degree. A higher degree, another angle, or another tiling may yield more, and the true shear modulus of the honeycomb is unknown. On the other side, the only upper bound is the trivial $\mu^{*}\le\mu$, so $P/Q\le S\le1$, and most of the remaining uncertainty lies on the upper side. Whether a nontrivial upper bound exists at zero bulk modulus and whether the construction has a three-dimensional analogue are the natural next questions.

\section*{Data availability}

The integer coefficients of the piecewise Airy potential of Result~\ref{res:2}, together with the exact values of $P$ and $Q$, are available from the author on request.

\bibliographystyle{unsrtnat}
\bibliography{references}

\begin{thebibliography}{18}
\providecommand{\natexlab}[1]{#1}
\providecommand{\url}[1]{\texttt{#1}}
\expandafter\ifx\csname urlstyle\endcsname\relax
  \providecommand{\doi}[1]{doi: #1}\else
  \providecommand{\doi}{doi: \begingroup \urlstyle{rm}\Url}\fi

\bibitem[Evans et~al.(1991)Evans, Nkansah, Hutchinson, and Rogers]{Evans1991}
K.~E. Evans, M.~A. Nkansah, I.~J. Hutchinson, and S.~C. Rogers.
\newblock Molecular network design.
\newblock \emph{Nature}, 353:\penalty0 124, 1991.
\newblock \doi{10.1038/353124a0}.

\bibitem[Greaves et~al.(2011)Greaves, Greer, Lakes, and Rouxel]{Greaves2011}
G.~N. Greaves, A.~L. Greer, R.~S. Lakes, and T.~Rouxel.
\newblock {Poisson's} ratio and modern materials.
\newblock \emph{Nature Materials}, 10:\penalty0 823--837, 2011.
\newblock \doi{10.1038/nmat3134}.

\bibitem[Lakes(1987)]{Lakes1987}
R.~Lakes.
\newblock Foam structures with a negative {Poisson's} ratio.
\newblock \emph{Science}, 235:\penalty0 1038--1040, 1987.
\newblock \doi{10.1126/science.235.4792.1038}.

\bibitem[Grima and Evans(2000)]{Grima2000}
J.~N. Grima and K.~E. Evans.
\newblock Auxetic behavior from rotating squares.
\newblock \emph{Journal of Materials Science Letters}, 19:\penalty0 1563--1565,
  2000.
\newblock \doi{10.1023/A:1006781224002}.

\bibitem[Prall and Lakes(1997)]{Prall1997}
D.~Prall and R.~S. Lakes.
\newblock Properties of a chiral honeycomb with a {Poisson's} ratio of $-1$.
\newblock \emph{International Journal of Mechanical Sciences}, 39:\penalty0
  305--314, 1997.
\newblock \doi{10.1016/S0020-7403(96)00025-2}.

\bibitem[Milton(1992)]{Milton1992}
G.~W. Milton.
\newblock Composite materials with {Poisson's} ratios close to $-1$.
\newblock \emph{Journal of the Mechanics and Physics of Solids}, 40:\penalty0
  1105--1137, 1992.
\newblock \doi{10.1016/0022-5096(92)90063-8}.

\bibitem[B{\"u}ckmann et~al.(2014)B{\"u}ckmann, Schittny, Thiel, Kadic, Milton,
  and Wegener]{Buckmann2014}
T.~B{\"u}ckmann, R.~Schittny, M.~Thiel, M.~Kadic, G.~W. Milton, and M.~Wegener.
\newblock On three-dimensional dilational elastic metamaterials.
\newblock \emph{New Journal of Physics}, 16:\penalty0 033032, 2014.
\newblock \doi{10.1088/1367-2630/16/3/033032}.

\bibitem[Milton(2021{\natexlab{a}})]{Milton2021open}
G.~W. Milton.
\newblock Some open problems in the theory of composites.
\newblock \emph{Philosophical Transactions of the Royal Society A:
  Mathematical, Physical and Engineering Sciences}, 379:\penalty0 20200115,
  2021{\natexlab{a}}.
\newblock \doi{10.1098/rsta.2020.0115}.

\bibitem[Avellaneda et~al.(1996)Avellaneda, Cherkaev, Gibiansky, Milton, and
  Rudelson]{Avellaneda1996}
M.~Avellaneda, A.~V. Cherkaev, L.~V. Gibiansky, G.~W. Milton, and M.~Rudelson.
\newblock A complete characterization of the possible bulk and shear moduli of
  planar polycrystals.
\newblock \emph{Journal of the Mechanics and Physics of Solids}, 44:\penalty0
  1179--1218, 1996.
\newblock \doi{10.1016/0022-5096(96)00018-X}.

\bibitem[Milton(2021{\natexlab{b}})]{Milton2021planar}
G.~W. Milton.
\newblock Planar polycrystals with extremal bulk and shear moduli.
\newblock \emph{Journal of the Mechanics and Physics of Solids}, 157:\penalty0
  104601, 2021{\natexlab{b}}.
\newblock \doi{10.1016/j.jmps.2021.104601}.

\bibitem[Hill(1963)]{Hill1963}
R.~Hill.
\newblock Elastic properties of reinforced solids: some theoretical principles.
\newblock \emph{Journal of the Mechanics and Physics of Solids}, 11:\penalty0
  357--372, 1963.
\newblock \doi{10.1016/0022-5096(63)90036-X}.

\bibitem[Cherkaev et~al.(1992)Cherkaev, Lurie, and Milton]{Cherkaev1992}
A.~V. Cherkaev, K.~A. Lurie, and G.~W. Milton.
\newblock Invariant properties of the stress in plane elasticity and
  equivalence classes of composites.
\newblock \emph{Proceedings of the Royal Society of London. Series A:
  Mathematical and Physical Sciences}, 438:\penalty0 519--529, 1992.
\newblock \doi{10.1098/rspa.1992.0123}.

\bibitem[Bacigalupo and Gambarotta(2020)]{Bacigalupo2020}
A.~Bacigalupo and L.~Gambarotta.
\newblock Chiral two-dimensional periodic blocky materials with elastic
  interfaces: auxetic and acoustic properties.
\newblock \emph{Extreme Mechanics Letters}, 39:\penalty0 100769, 2020.
\newblock \doi{10.1016/j.eml.2020.100769}.

\bibitem[Ravirala et~al.(2007)Ravirala, Alderson, and Alderson]{Ravirala2007}
N.~Ravirala, A.~Alderson, and K.~L. Alderson.
\newblock Interlocking hexagons model for auxetic behaviour.
\newblock \emph{Journal of Materials Science}, 42:\penalty0 7433--7445, 2007.
\newblock \doi{10.1007/s10853-007-1583-0}.

\bibitem[Acuna et~al.(2022)Acuna, Guti{\'e}rrez, Silva, Palza, Nunez, and
  D{\"u}ring]{Acuna2022}
D.~Acuna, F.~Guti{\'e}rrez, R.~Silva, H.~Palza, A.~S. Nunez, and G.~D{\"u}ring.
\newblock A three step recipe for designing auxetic materials on demand.
\newblock \emph{Communications Physics}, 5:\penalty0 113, 2022.
\newblock \doi{10.1038/s42005-022-00876-5}.

\bibitem[Sigmund(2000)]{Sigmund2000}
O.~Sigmund.
\newblock A new class of extremal composites.
\newblock \emph{Journal of the Mechanics and Physics of Solids}, 48:\penalty0
  397--428, 2000.
\newblock \doi{10.1016/S0022-5096(99)00034-4}.

\bibitem[Ostanin et~al.(2018)Ostanin, Ovchinnikov, Tozoni, and
  Zorin]{Ostanin2018}
I.~Ostanin, G.~Ovchinnikov, D.~C. Tozoni, and D.~Zorin.
\newblock A parametric class of composites with a large achievable range of
  effective elastic properties.
\newblock \emph{Journal of the Mechanics and Physics of Solids}, 118:\penalty0
  204--217, 2018.
\newblock \doi{10.1016/j.jmps.2018.05.018}.

\bibitem[Milton et~al.(2017)Milton, Briane, and Harutyunyan]{Milton2017}
G.~W. Milton, M.~Briane, and D.~Harutyunyan.
\newblock On the possible effective elasticity tensors of 2-dimensional and
  3-dimensional printed materials.
\newblock \emph{Mathematics and Mechanics of Complex Systems}, 5:\penalty0
  41--94, 2017.
\newblock \doi{10.2140/memocs.2017.5.41}.

\end{thebibliography}

\clearpage
\appendix
\makeatletter
\renewcommand{\@seccntformat}[1]{\csname pre#1\endcsname\csname the#1\endcsname\csname post#1\endcsname\quad}
\makeatother
\newcommand{\presection}{Appendix~}
\newcommand{\postsection}{.}
\newcommand{\presubsection}{}
\newcommand{\postsubsection}{}

\section{The thickened tiling}\label{sec:A}

This appendix constructs the geometry used in Section~\ref{sec:2.3} and provides the block motion and interface tractions required in Appendix~\ref{sec:B} for the proof of Result~\ref{res:3}. The geometric construction is valid for every $\theta>0$ and $0^{\circ}<\gamma<90^{\circ}$.

\subsection{Geometry}\label{sec:A.1}

Let $\Omega$ be the complete hexagonal cell chosen in Section~\ref{sec:2.1}, centered at the origin, with vertices $V_{k}=(\cos(60^{\circ}k), \sin(60^{\circ}k))$, where the indices are taken modulo six. These Cartesian vertices have the oblique coordinates $v_{k}$ introduced in Section~\ref{sec:2.2}. Edge $k$ runs from $V_{k}$ to $V_{k+1}$, with a unit tangent $\tau_{k}=V_{k+1}-V_{k}$ and an outward unit normal $n_{k}=(V_{k}+V_{k+1})/\sqrt{3}$, so $\tau_{k}=Jn_{k}$. We use the directions $\hat{a}_{k}$ and $\hat{t}_{k}$ defined by \eqref{eq:5}, so $\hat{t}_{k}\cdot n_{k}=-\sin\gamma$ and $\hat{t}_{k}\cdot\tau_{k}=\cos\gamma$.

We write $c=\cot\gamma$, let $\Lambda$ be the triangular lattice generated by the nearest-neighbor vectors $\sqrt{3}\, n_{k}$, and set
\begin{equation}\label{eq:16}
  A_{\theta}=I+\frac{\theta}{\sqrt{3}}(I-cJ),\quad w_{k}=\theta(I-cJ)n_{k}=\theta n_{k}-\theta c\tau_{k}=\ell\hat{t}_{k},\quad \ell=-\frac{\theta}{\sin\gamma}<0.
\end{equation}
The new block centers form the lattice $\Lambda_{\theta}=A_{\theta}\Lambda$. Since $n_{k}$ and $\tau_{k}$ are orthonormal, the vectors $w_{k}$ have equal length $\theta\sqrt{1+c^{2}}=\theta/\sin\gamma=|\ell|$ and successive directions differ by $60^{\circ}$. Here $\ell$ is signed; the distance between corresponding points on the facing block edges is $|\ell|$.

The thickened tiling $H_{\gamma,\theta}$ consists of blocks $x+\Omega$, with $x\in\Lambda_{\theta}$; strips
\begin{equation}\label{eq:17}
  \mathcal{S}_{x,k}=\{x+V_{k}+s\tau_{k}+\eta w_{k}:\: 0\le s\le1,\: 0\le\eta\le1\};
\end{equation}
and triangles $T_{x,j}=\conv\{x+V_{j},\, x+V_{j}+w_{j},\, x+V_{j}+w_{j-1}\}$ at the block vertices (Fig.~\ref{fig:2}). We call $k$ the class of the strip $\mathcal{S}_{x,k}$. Each strip has normal thickness $\theta$ and will contain the plates of Appendix~\ref{sec:B.1}; the triangles remain void. We write $\Omega_{\theta}$ for a periodic cell of $\Lambda_{\theta}$.

\subsection{The exact tiling}\label{sec:A.2}

\textbf{Facing edges.} Consider the block $A$ at the origin and its neighbor $B$ centered at $x'=A_{\theta}(\sqrt{3}\, n_{k})=\sqrt{3}\, n_{k}+w_{k}$. Parametrizing their facing edges in the common direction $\tau_{k}$ (Fig.~\ref{fig:2}b) yields
\begin{equation}\label{eq:18}
  \begin{aligned} X_{A}(s) &=V_{k}+s\tau_{k}, \\ X_{B}(s) &=x'-V_{k+1}+s\tau_{k}=X_{A}(s)+w_{k},\qquad 0\le s\le1. \end{aligned}
\end{equation}
The equality follows from $\sqrt{3}\, n_{k}=V_{k}+V_{k+1}$. Therefore, the strip $\mathcal{S}_{0,k}$ is the union of the line segments joining $X_{A}(s)$ to $X_{B}(s)$, all parallel to $\hat{t}_{k}$. On edge $k+3$ of $B$, the corresponding local edge parameter is $1-s$.

\textbf{A common strip.} Since $V_{k+3}=-V_{k}$, $\tau_{k+3}=-\tau_{k}$ and $w_{k+3}=-w_{k}$, the parametrization of $\mathcal{S}_{x',k+3}$ becomes that of $\mathcal{S}_{0,k}$ under
\begin{equation}\label{eq:19}
  (s,\eta)\mapsto(1-s,1-\eta).
\end{equation}
The two neighboring blocks define the same strip. Translation extends the conclusion to every pair of neighbors. A plate pattern prescribed identically from both blocks is well defined if it is invariant under \eqref{eq:19}.

\textbf{Junction triangles and area.} The identity $V_{j-1}+V_{j+1}=V_{j}$ implies $n_{j-1}+n_{j+1}=n_{j}$ and, since $w_{k}$ is linear in $n_{k}$ by \eqref{eq:16}, $w_{j-1}+w_{j+1}=w_{j}$. Therefore, all three sides of $T_{0,j}$ have length $|\ell|=\theta/\sin\gamma$, and the triangle is equilateral (Fig.~\ref{fig:2}c). Its vertices are corners of the block at the origin and of the neighbors across edges $j$ and $j-1$. Two sides are lateral boundaries of $\mathcal{S}_{0,j}$ and $\mathcal{S}_{0,j-1}$; the third is the segment $V_{j}+w_{j-1}+\eta w_{j+1}$, $0\le\eta\le1$, which is the $s=1$ boundary of the strip of class $j+1$ between those neighbors. Describing the junction from either neighbor leads to the same triangle.

Each strip is associated with two block edges and each triangle with three block vertices. Therefore, one primitive cell carries one block, three strips, and two triangles. Their areas are $3\sqrt{3}/2$, $\theta$ per strip and $\sqrt{3}\, \theta^{2}(1+c^{2})/4$ per triangle. Since $\det A_{\theta}=(1+\theta/\sqrt{3})^{2}+\theta^{2}c^{2}/3$, their sum is exactly the cell area:
\begin{equation}\label{eq:20}
  |\Omega_{\theta}|=\det(A_{\theta})|\Omega|=\frac{3\sqrt{3}}{2}+3\theta+\frac{\sqrt{3}}{2}\theta^{2}(1+c^{2}).
\end{equation}
\textbf{Global tiling.} The prescribed edge identifications pair every block edge with a strip face and every lateral strip boundary with a triangle side. The two pieces in each pair lie on opposite sides of their common edge. At a block vertex $V_{j}$, the four pieces prescribed by these identifications occur in the cyclic order: block, strip $\mathcal{S}_{0,j-1}$, triangle $T_{0,j}$ and strip $\mathcal{S}_{0,j}$. Their wedges lie between the consecutive rays $\tau_{j}$, $-\tau_{j-1}$, $w_{j-1}$, $w_{j}$ and $\tau_{j}$, and have angles $120^{\circ}$, $\gamma$, $60^{\circ}$ and $180^{\circ}-\gamma$. These angles are positive and sum to $360^{\circ}$. Every corner of a strip or triangle is a block corner, so this checks all vertices.

We glue the pieces modulo $\Lambda_{\theta}$ along their prescribed paired edges. The resulting compact surface maps to the torus $\mathbb{R}^{2}/\Lambda_{\theta}$. The edge pairing and the full angle at every vertex make this map locally one-to-one, so it is a finite covering. Its total area equals the area of the torus by \eqref{eq:20}, forcing the covering degree to be one. Therefore, the blocks, strips, and triangles tile the plane with disjoint interiors for every stated $\theta$ and $\gamma$.

\subsection{Symmetry and the rigid block motion}\label{sec:A.3}

The map $A_{\theta}=(1+\theta/\sqrt{3})I-(\theta c/\sqrt{3})J$ has the form $aI+bJ$ and is therefore a similarity that commutes with rotations, so $\Lambda_{\theta}$ is a rotated and rescaled copy of $\Lambda$. Consequently, rotation through $60^{\circ}$ about any block center permutes the blocks, strips, and triangles.

We give each block the rigid displacement $u_{x}(X)=ex+\omega J(X-x)$, as in \eqref{eq:6}. We choose $\omega$ so that the relative displacement of corresponding plate endpoints has no component along the plate axis $\hat{t}_{k}$, so that the plates bend without stretching. For the pair in \eqref{eq:18}, $X_{B}-X_{A}=w_{k}=\ell\hat{t}_{k}$, and $(J\hat{t}_{k})\cdot\hat{t}_{k}=0$ removes the rotation term $\omega J(X_{B}-X_{A})$ from this component. Using $x'=(\sqrt{3}+\theta)n_{k}-\theta c\tau_{k}$ and $c\sin\gamma=\cos\gamma$ yields
\begin{equation}\label{eq:21}
  [u_{x'}(X_{B}(s))-u_{0}(X_{A}(s))]\cdot\hat{t}_{k}=-\frac{e(\sqrt{3}\sin^{2}\gamma+\theta)}{\sin\gamma}-\omega\sqrt{3}\cos\gamma.
\end{equation}
This expression is independent of $s$ and of the edge class, and vanishes when
\begin{equation}\label{eq:22}
  \omega=-e\, \frac{\sin^{2}\gamma+\theta/\sqrt{3}}{\sin\gamma\cos\gamma}.
\end{equation}
As $\theta\to0$, this reduces to the ideal relation $\omega=-e\tan\gamma$ in \eqref{eq:8}. For every $r\in\Lambda_{\theta}$, the block displacement satisfies $u_{x+r}(X+r)-u_{x}(X)=er$. Therefore, $u-eX$ is periodic on the blocks and the macroscopic strain is $eI$. Appendix~\ref{sec:B.2} extends these motions through the plates using an incompressible bending field whose energy tends to zero as the plates become finer at fixed $\theta$.

\subsection{The ideal field on the thickened tiling}\label{sec:A.4}

Let $\sigma_{0}$ be an admissible stress field for $H_{\gamma}$ as in Section~\ref{sec:2.2}, with Lipschitz traction along each interface. By the sliding condition, $\sigma_{0}n=q\hat{t}$, where $q=(\sigma_{0}n)\cdot\hat{t}$. We translate the stress in each original block to the corresponding block of $H_{\gamma,\theta}$; these copies are periodic with respect to $\Lambda_{\theta}$.

For the pair in \eqref{eq:18}, block $A$ is unchanged, and block $B$ is translated by $w_{k}$. Therefore, $X_{B}(s)-w_{k}=X_{A}(s)$: the two new facing points pull back to the same spatial location on the original shared interface. Traction continuity there yields $q(s)\hat{t}_{k}$ on the new face of $A$, with outward normal $n_{k}$, and $-q(s)\hat{t}_{k}$ on the new face of $B$, with outward normal $-n_{k}$. These are the tractions to be transmitted through the plates in Appendix~\ref{sec:B.3}.

We write
\begin{equation}\label{eq:23}
  M_{0}=\int_{\Omega} \sigma_{0}\, dX,\qquad E_{0}=\frac{1}{4\mu}\int_{\Omega} |\dev\sigma_{0}|^{2}\, dX
\end{equation}
for the integrated stress and complementary energy over the original hexagonal cell. With $\Sigma=M_{0}/|\Omega|$ and $W_{c}=E_{0}/|\Omega|$, the quotient in \eqref{eq:10} is $|M_{0}|^{2}/(4\mu|\Omega|E_{0})$ when $E_{0}>0$. Also, $\tr M_{0}=0$ by the argument following \eqref{eq:9} in Section~\ref{sec:2.2}. These quantities supply the reference stress and energy for the estimates in Appendix~\ref{sec:B}.

\section{The solid--void structures}\label{sec:B}

This appendix constructs the solid--void mixtures used in Section~\ref{sec:2.3} and proves Result~\ref{res:3}. We fix $0^{\circ}<\gamma<90^{\circ}$ and $p\in(0,1)$, and let $\sigma_{0}$ be an admissible stress field for $H_{\gamma}$ with Lipschitz traction on each interface. We first let $N\to\infty$ at fixed $\theta>0$, and then let $\theta\to0$. Constants denoted by $C$, $C_{1}$ or $D_{\theta}$ may depend on $\mu,\gamma,p,\theta$ and $\sigma_{0}$, but not on $N$. The constants $C_{m}$ and $C_{e}$ defined below are independent of both $\theta$ and $N$.

\subsection{Geometry}\label{sec:B.1}

Let $N\ge3$ be an integer. In each strip $\mathcal{S}_{x,k}$ of \eqref{eq:17}, we use the coordinates $(s,\eta)$ to define two solid end slabs, $0\le\eta\le1/N$ and $1-1/N\le\eta\le1$, spanning the full block faces. For $j=0,\ldots,N-1$, write $I_{j}=[j/N,(j+1)/N]$ and $s_{j}=(2j+1)/(2N)$. Between the slabs, place the plates
\begin{equation}\label{eq:24}
  P_{j}=\left\{ x+V_{k}+s\tau_{k}+\eta w_{k}:\: |s-s_{j}|\le\frac{p}{2N},\quad \frac{1}{N}\le\eta\le1-\frac{1}{N} \right\}.
\end{equation}
The solid consists of the blocks, end slabs, and plates. The remaining strip regions and the junction triangles are void (Fig.~\ref{fig:3}). Each plate runs parallel to $\hat{t}_{k}$ and occupies the fraction $p$ of its face interval $I_{j}$.

The reversal $(s,\eta)\mapsto(1-s,1-\eta)$ in \eqref{eq:19} permutes the plates and exchanges the two end slabs, so the construction is well defined from either neighboring block. Rotation through $60^{\circ}$ preserves the strip coordinates and permutes the six strip classes. Therefore, $H_{\gamma,\theta}^{N}$ is a periodic solid--void mixture with sixfold symmetry, and its effective tensor $C_{N}^{*}=C^{*}(H_{\gamma,\theta}^{N})$ is isotropic.

\subsection{Vanishing bulk modulus}\label{sec:B.2}

We prescribe the macroscopic strain $E=eI$ and move the blocks by the rigid displacements $u_{x}$ of Appendix~\ref{sec:A.3}, with $\omega$ given by \eqref{eq:22}, and let each end slab move with the block it touches. For neighboring blocks, define $\Delta=u_{x'}(X)-u_{x}(X)$, evaluating both affine rigid-motion formulas at the same spatial point $X$. This difference is constant. Its component along $\hat{t}_{k}$ equals that of the endpoint difference in \eqref{eq:21}, because the additional term $\omega Jw_{k}$ is perpendicular to $\hat{t}_{k}$. Therefore, $\Delta=\delta\hat{a}_{k}$, where $\delta$ is proportional to $e$ by \eqref{eq:22}, with $|\delta|\le C|e|$.

For plate $j$, take orthonormal coordinates $(\xi,\zeta)$ along $(\hat{t}_{k},\hat{a}_{k})$, with origin at $x+V_{k}+s_{j}\tau_{k}$. Its width is $h_{p}=p\sin\gamma/N$, and $\xi=\zeta\cot\gamma+\eta\ell$, where $\ell<0$ is defined in \eqref{eq:16}. The two oblique end faces have a common axial gap $[\xi_{-},\xi_{+}]$, with the $x'$ end at smaller $\xi$ and the $x$ end at larger $\xi$. The gap length is $L=|\ell|(1-2/N)-p\cos\gamma/N$. At fixed $\theta$, it tends to $|\ell|>0$, so $L\ge|\ell|/2$ for sufficiently large $N$.

We choose a smooth function $\chi$, obtained by rescaling a fixed profile, that equals $\delta$ for $\xi\le\xi_{-}$ and $0$ for $\xi\ge\xi_{+}$. In the plate set
\begin{equation}\label{eq:25}
  u=u_{x}-\zeta\chi'(\xi)\hat{t}_{k}+\left(\chi(\xi)+\frac{1}{2}\zeta^{2}\chi''(\xi) \right)\hat{a}_{k}.
\end{equation}
The divergence of the axial correction is $-\zeta\chi''$, while that of the transverse correction is $\zeta\chi''$, so the field is exactly incompressible. On the full oblique end faces, $\chi$ is constant and its derivatives vanish. The field matches $u_{x}$ at one slab and $u_{x}+\Delta=u_{x'}$ at the other. We choose these extensions in one period and repeat them by lattice translations; then $u-eX$ is periodic throughout the solid.

The rescaled profile satisfies $|\chi''|\le C|\delta|/L^{2}$ and $|\chi'''|\le C|\delta|/L^{3}$. Since $|\zeta|\le h_{p}/2$ and $h_{p}/L$ is bounded, the strain is $O(h_{p}|\delta|/L^{2})$. Its support lies in a transition region of area at most $h_{p}L$, yielding
\begin{equation}\label{eq:26}
  \int_{P_{j}} 2\mu|\epsilon(u)|^{2}\, dX\le C\mu\frac{h_{p}^{3}\delta^{2}}{L^{3}}\le C\mu\frac{\delta^{2}}{N^{3}}.
\end{equation}
The blocks and end slabs have zero strain. For sufficiently large $N$, summing over the $3N$ plates per cell and using the cell-energy principle \eqref{eq:4} yields
\begin{equation}\label{eq:27}
  \kappa^{*}(C_{N}^{*})\le C_{1}N^{-2}.
\end{equation}
For any deviatoric strain $E$, the affine displacement $u(X)=EX$ is incompressible and shows that  $\mu^{*}(C_{N}^{*})\le\mu$. Adding such a displacement to the dilation field provides a finite-energy trial field for every symmetric macroscopic strain. The isotropic tensors $C_{N}^{*}$ are finite and bounded as $N\to\infty$, and their bulk moduli tend to zero.

\subsection{An admissible stress field}\label{sec:B.3}

We translate $\sigma_{0}$ into each block as in Appendix~\ref{sec:A.4}. In plate $P_{j}$ of a class-$k$ strip, choose the axial stress $\sigma=f_{j}(s)\hat{t}_{k}\otimes\hat{t}_{k}$, with $f_{j}(s)=a_{j}+b_{j}N(s-s_{j})$. It is divergence-free because $s$ is constant along $\hat{t}_{k}$, and its long sides are traction-free.

We divide each end slab into the cells $s\in I_{j}$. We write $\alpha=\hat{t}_{k}\cdot n_{k}=-\sin\gamma$. On the cell next to block $x$, the prescribed outward traction is $-q(s)\hat{t}_{k}$ on the block face, $\alpha f_{j}(s)\hat{t}_{k}$ on the plate contact, and zero on the two flanking gaps and the two lateral cell sides. At the opposite end slab, both nonzero tractions reverse sign, with the same parameter $s$, by Appendix~\ref{sec:A.4}.

All forces are parallel to $\hat{t}_{k}$, so force and moment balance reduce to two scalar conditions. The plate is centered in $I_{j}$: its constant stress coefficient controls the resultant force, and its linear coefficient controls the moment. The face offset $w_{k}/N$ is parallel to the forces and adds no moment. Both conditions are satisfied by
\begin{equation}\label{eq:28}
  a_{j}=\frac{N}{p\alpha}\int_{I_{j}} q(s)\, ds,\qquad b_{j}=\frac{12N^{2}}{p^{3}\alpha}\int_{I_{j}} (s-s_{j})q(s)\, ds.
\end{equation}
The same coefficients balance the cell at the opposite end of the plate.

Let $Q_{\infty}$ bound $|q|$ and let $L_{q}$ bound its Lipschitz constant on all six block faces. Equation \eqref{eq:28} yields $|a_{j}|\le Q_{\infty}/(p|\alpha|)$. Subtracting $q(s_{j})$ in the integral for $b_{j}$ and using $\int_{I_{j}} (s-s_{j})^{2}\, ds=1/(12N^{3})$ yields $|b_{j}|\le L_{q}/(p^{3}|\alpha|N)$. Since $|N(s-s_{j})|\le p/2$ on the plate,
\begin{equation}\label{eq:29}
  |f_{j}(s)|\le F:=\frac{Q_{\infty}}{p|\alpha|}+\frac{L_{q}}{2p^{2}|\alpha|}.
\end{equation}
This bound is independent of both $\theta$ and $N$.

Each end-slab cell is a translated copy, scaled by $N^{-1}$, of a fixed parallelogram $Y$ with side vectors $\tau_{k}$ and $w_{k}$. At fixed $\theta$, this is a bounded Lipschitz domain. The prescribed tractions have zero total force and moment by \eqref{eq:28}. An auxiliary coercive linear-elasticity Neumann problem on $Y$, modulo rigid motions, therefore supplies a symmetric equilibrium stress $\sigma_{Y}$. Korn's inequality, the trace inequality and the Lax--Milgram lemma yield $\lVert\sigma_{Y}\rVert_{L^{2}(Y)}\le C\lVert g_{Y}\rVert_{L^{2}(\partial Y)}$, where $g_{Y}$ denotes the boundary data pulled back to $Y$. Only the resulting stress is used; the auxiliary material law need not be that of the incompressible constituent.

The pulled-back tractions are uniformly bounded by $\max(Q_{\infty},|\alpha|F)$. Scaling the stress back to a cell preserves its traction and multiplies its $L^{2}$ norm by $N^{-1}$. There are $6N$ cells per period, with total area $6\theta/N$. Summation and the Cauchy--Schwarz inequality yield
\begin{equation}\label{eq:30}
  \int_{\text{end slabs}} |\sigma|^{2}\, dX\le\frac{C}{N},\qquad \int_{\text{end slabs}} |\sigma|\, dX\le\frac{C}{N}.
\end{equation}
The constant may depend on $\theta$ through the shape of $Y$. Zero traction on the lateral cell sides permits the cell stresses to be joined, while the prescribed face tractions match the block and plate stresses. Repeating the construction periodically produces a symmetric, divergence-free stress with continuous internal traction and zero traction on every solid--void boundary.

We extend this stress by zero into the void and use $M_{\theta,N}$ and $E_{\theta,N}$ as in Section~\ref{sec:2.3}. The blocks contribute exactly $M_{0}$ and $E_{0}$ from \eqref{eq:23}. For a plate stress, $|\sigma|=|f_{j}|$ and $|\dev\sigma|^{2}=f_{j}^{2}/2$. The plates have total area $3p(1-2/N)\theta\le3p\theta$ per period. Together with \eqref{eq:29}--\eqref{eq:30}, these facts prove \eqref{eq:13} with $C_{m}=3pF$ and $C_{e}=3pF^{2}/(8\mu)$. These constants are independent of $\theta$ and $N$; the terms $D_{\theta}N^{-1}$ account for the end slabs.

Finally, put $\Sigma=M_{\theta,N}/|\Omega_{\theta}|$ and $W=E_{\theta,N}/|\Omega_{\theta}|$. Restricting \eqref{eq:9} to deviatoric macroscopic strains yields $W\ge\sup_{\tr E=0}(\dev\Sigma\cdot E-\mu^{*}(C_{N}^{*})|E|^{2})$. If $\dev\Sigma\ne0$, finiteness forces $\mu^{*}(C_{N}^{*})>0$, and the supremum is $|\dev\Sigma|^{2}/(4\mu^{*}(C_{N}^{*}))$. This proves \eqref{eq:14}, without any assumption on $\tr\Sigma$.

\subsection{Passage to the limit}\label{sec:B.4}

We fix $\theta>0$. The bounded tensors $C_{N}^{*}$ have convergent subsequences as $N\to\infty$. Every limit $C_{\infty,\theta}$ is isotropic and has zero bulk modulus by \ref{sec:B.2}. It also belongs to $\mathcal{G}$, because each $C_{N}^{*}$ is the finite effective tensor of a periodic mixture of the constituent solid and void, and $\mathcal{G}$ is closed.

We assume $M_{0}\ne0$ and $E_{0}>0$, as holds for the stress field of Result~\ref{res:2}. Since $\dev M_{0}=M_{0}$ by Appendix~\ref{sec:A.4} and the deviatoric projection is a contraction, \eqref{eq:13} implies $|\dev M_{\theta,N}|\ge|M_{0}|-C_{m}\theta-D_{\theta}/N$. We choose $\theta$ so that $C_{m}\theta<|M_{0}|$; the right-hand side is then positive for all sufficiently large $N$. Applying \eqref{eq:13}--\eqref{eq:14} and passing along a convergent subsequence yields
\begin{equation}\label{eq:31}
  \frac{\mu^{*}(C_{\infty,\theta})}{\mu}\ge\frac{(|M_{0}|-C_{m}\theta)^{2}}{4\mu|\Omega_{\theta}|(E_{0}+C_{e}\theta)}.
\end{equation}
Here the shear modulus depends continuously on the isotropic tensor. The bound applies to every subsequential limit, so convergence of the full sequence is unnecessary.

Each $C_{\infty,\theta}$ is admissible in the definition \eqref{eq:2} of $S$. Consequently, $S$ is at least the right-hand side of \eqref{eq:31} for every sufficiently small $\theta$. By \eqref{eq:20}, $|\Omega_{\theta}|\to|\Omega|$ as $\theta\to0$, and that right-hand side tends to $|M_{0}|^{2}/(4\mu|\Omega|E_{0})$, the quotient in \eqref{eq:10}. We choose $p=1/2$ and the field of Result~\ref{res:2} at $\tan\gamma=100/\sqrt{3}$. Being polynomial, its interface tractions are Lipschitz, and its quotient equals $P/Q$. Therefore $S\ge P/Q>0.8528>4/5$, proving Result~\ref{res:3}.

Since $P/Q>4/5$, the right-hand side of \eqref{eq:31} exceeds $4/5$ for sufficiently small fixed $\theta$. The estimates \eqref{eq:13}--\eqref{eq:14} before passage to the limit then show that $\mu^{*}(C_{N}^{*})/\mu>4/5$ for all sufficiently large finite $N$. Together with \eqref{eq:27}, this implies that the effective Poisson's ratio tends to $-1$ as $N\to\infty$. The argument requires no bound on $D_{\theta}$ uniform in $\theta$, because the limit in $N$ is taken first.

\end{document}